\documentclass[sigconf]{acmart}

\usepackage[most]{tcolorbox}
\AtBeginDocument{%
  }

\acmConference[CONF 2025]{1st Super Important Conference on Software Engineering}{December 24,
  2025}{New York City, USA}

\begin{document}

%\title{Exploring the Performance of Autistic Individuals in Software Engineering Activities - A Mixed-Methods Study}
\title{Investigating the Experience of Autistic Individuals in Software Engineering}

\author{Madalena Sasportes}
\email{madasasportes@gmail.com}
\orcid{0009-0002-8282-8284}
\affiliation{%
  \institution{NOVA SST, NOVA LINCS}
  \city{Lisbon}
  \country{Portugal}
}

\author{Grischa Liebel}
\email{grischal@ru.is}
\orcid{0000-0002-3884-815X}
\affiliation{%
  \institution{Reykjavik University}
  \city{Reykjavik}
  \country{Iceland}}

\author{Miguel Goulão}
\orcid{0000-0002-5356-5203}
\email{mgoul@fct.unl.pt}
\affiliation{%
  \institution{NOVA SST, NOVA LINCS}
  \city{Lisbon}
  \country{Portugal}
}

\renewcommand{\shortauthors}{Sasportes, et al.}

\begin{abstract}
\textit{Context:} Autism spectrum disorder (ASD) leads to various issues in the everyday life of autistic individuals, often resulting in unemployment and mental health problems.
To improve the inclusion of autistic adults, existing studies have highlighted the strengths these individuals possess in comparison to non-autistic individuals, e.g., high attention to detail or excellent logical reasoning skills.
If fostered, these strengths could be valuable in software engineering activities, such for identifying specific kinds of bugs in code.
However, existing work in SE has primarily studied the challenges of autistic individuals and possible accommodations, with little attention their strengths.

\textit{Objective:} 
Our goal is to analyse the experiences of autistic individuals in software engineering activities, such as code reviews, with a particular emphasis on strengths.
%This study explores the experiences of individuals with ASD in diverse software engineering contexts, aiming to create a more inclusive environment for neurodivergent software engineers. We align with the United Nations’ Sustainable Development Goals 4 (Ensure inclusive and equitable quality education) and 10 (Reduce inequality within and among countries) to enhance the representation and success of individuals with ASD in higher education and the software engineering industry.\\

\textit{Methods:} This study combines Social-Technical Grounded Theory through semi-structured interviews with 16 autistic software engineers and a survey with 49 respondents, including 5 autistic participants.
We compare the emerging themes with the theory by Gama et al. on the Effect of Neurodivergent Cognitive Dysfunctions in
Software Engineering Performance.

\textit{Results:} Our results suggest that autistic software engineers are often skilled in logical thinking, attention to detail, and hyperfocus in programming; and they enjoy learning new programming languages and programming-related technologies. 
Confirming previous work, they tend to prefer written communication and remote work.
Finally, we report a high comfort level in interacting with AI-based systems.

\textit{Conclusions:} Our findings extend existing work by providing further evidence on the strengths of autistic software engineers.

%Autistic individuals offer unique skills and traits in software development and code reviews, differing from neurotypical individuals. With specific accommodations, most challenges they face can be mitigated. Companies should actively consider them in job applications due to their significant contributions.\\

\end{abstract}
\renewcommand{\shortauthors}{Anonymous, et al.}




\maketitle

\section{Introduction}
%\grischa{Refer to 2025 GT, and use this as a "validation"/comparison.}
%\grischa{Compile dataset.}
%Neurodivergent individuals vary in their neurocognitive function compared to neurotypical individuals \cite{todo}. 
Autism spectrum disorder (ASD) is a neurological disorder that leads to various issues in the everyday life of autistic individuals, such as difficulties with communication with non-autistic individuals~\cite{Williams2021Mutual}, teamwork~\cite{Holmes2020dark}, or forming lasting friendships \cite{Sperry2005Perceptions}.
As a result, 50-75\% of autistic adults are unemployed \cite{hendricks_2010_employment}.

%Specifically, many people with Autism Spectrum Disorder (ASD), which is part of the neurodiversity term, are more likely to pursue education and careers in STEM (science, technology, engineering, and mathematics) subjects compared to their peers \cite{XinWei}.
%In the United States, an estimated 49,000 autistic students graduated high school in 2015 \cite{wei_2015_the}, with many expected to pursue higher education. As promising as this sounds, research \cite{allen_2021_what, viezel_2020_collegebased} has indicated that these students have lower graduation rates and lower rates of post-graduation employment, compared to those without \emph{ASD}. In 2020, it was estimated that the post-secondary graduation rate of students with \emph{ASD} is only 38.8\% \cite{viezel_2020_collegebased}. Therefore, it is highly important to do research about individuals with higher levels of functioning as they exit high school. 

To improve inclusion of autistic adults, existing work has pointed out the many strengths these individuals posses in comparison to non-autistic individuals, e.g., high attention to detail, fine motor skills, excellent logical reasoning, and improved concentration~\cite{lorenz_2014_aspergers}.
In addition to accommodating for challenges autistic individuals have, these strengths can form a competitive advantage in many industries.
Specifically, various of the highlighted strengths could be valuable in software engineering (SE) activities, such as for identifying specific kinds of bugs in code or other debugging activities \cite{costello}.

Previous work in SE has studied the challenges of autistic individuals \cite{Morris,Marquez2024IST,Gama2023Understanding,Gama2025ICSE-SEIS}.
These works provide a comprehensive overview of challenges experienced by autistic software engineers, as well as potential explanations and accommodations.
Recently, this resulted in a theory~\cite{Gama2025ICSE-SEIS} that aims to explain how cognitive dysfunctions and communication issues lead to a stress response, thus affecting SE performance; as well as how accommodations and individual experiences can moderate this effect.
However, existing work focuses on generic workplace challenges, without a dedicated focus to SE activities.
Additionally, existing work covers primarily challenges.
%, not paying due attention to reported strengths.

To address this gap, in this study we investigate SE-related challenges and strengths reported by autistic software engineers.
We answer the following research question (RQ):
%\grischa{Re-visit RQs!}
\begin{enumerate}
%    \item[RQ1] How can we help individuals with ASD thrive in the field of Software Engineering?
    \item[RQ] What strengths and challenges do autistic individuals exhibit in SE-related activities?
%    \item[RQ2] How can software development teams and companies leverage the skills of individuals with \emph{ASD}?
\end{enumerate}
To answer this question, we conducted a qualitative study following the Socio-Technical Grounded Theory (STGT) method \cite{hoda_2022_sociotechnical}.
We collected data through 16 interviews with 11 software engineers and 5 SE graduate students.
To support the emerging themes, we followed up with a survey among 49 individuals.
We then systematically compare the findings with the theory proposed by Gama et al.~\cite{Gama2025ICSE-SEIS} and the challenges reported by Morris et al.~\cite{Morris}.
We confirm several of the findings in existing work, thus strengthening existing theory.
Additionally, we extend Gama et al.'s work in terms of SE-related challenges and strengths.
In particular, we find that communication challenges extend into code understanding.
We hypothesise that the \emph{double empathy problem} theory \cite{milton2012double,milton2022double} extends to code, i.e., that autistic individuals have difficulties understanding code and abstractions of neurotypical individuals, and vice versa.
As a result, we believe that team composition is an important moderating factor for the performance of autistic software engineers, thus adding it to the theory.
Finally, we add the strengths dimension to the existing theory, i.e., that strengths of autistic individuals likewise (positively) affect their SE performance.

\section{Background and Related Work}
\subsection{Neurodiversity}
Neurodiversity describes the idea that people experience and interact with the world around them in many different ways \cite{singer1998}\footnote{The term neurodiversity is usually attributed to Judy Singer. However, a recent article points out that this term pre-dates Singer's thesis and was most likely developed collectively by a community \cite{botha2024neurodiversity}.}. It acknowledges that there is no one ``right'' way of thinking, learning, and behaving, and differences should not be considered as shortcomings \cite{Pisano}. 
Neurodiversity encompasses individuals with various diagnosed neurological conditions, such as ASD, ADHD and dyslexia. 
However, the neurodiversity movement highlights that cognitive diversity should be considered natural variation that is to be respected and accommodated for, rather than a disease that needs cure.
An estimated 15\% to 20\% of the world population are neurodivergent~\cite{doyle2020neurodiversity}.
In contrast to neurodivergent individuals, \textit{neurotypical} describes individuals that are not neurodivergent. 

\subsection{Autism Spectrum Disorder (ASD)}
ASD is a lifelong neurodevelopmental condition characterised by particular cognitive styles, communication behaviours, social interactions, and repetitive behaviours~\cite{american2013diagnostic}. 
Because ASD occurs on a spectrum, it presents itself in different degrees of severity. The \emph{DSM-5} classifies it into three different severity levels. These levels are determined by the level of support an individual requires to function in daily life, where level 1 includes individuals that experience only subtle symptoms that do not interfere with their ability to function at school, work or in relationships; level 2 describes individuals that need more support, such as speech therapy to improve language skills or interventions aimed at social skills and reducing repetitive behaviours; level 3 describes individuals that depend on support for most tasks \cite{american2013diagnostic}.
While originally a separate diagnosis, \textit{Asperger Syndrome} is now classified as level 1 or level 2 ASD.
ASD commonly co-occurs, or is \emph{co-morbid} with other cognitive disorders. For instance, evidence indicates that 50\% to 70\% of individuals with ASD also have ADHD~\cite{hours2022asd}.

The prevalence of ASD has risen in past years~\cite{Talantseva2023The}, although a significant portion of it may be explained by improvements in diagnosis practices, awareness and service access~\cite{OSharkey2024Trends}. According to the Center for Disease Control and Prevention~\cite{cdc_prevalence}, the prevalence of ASD is approximately 1 in 31 children aged 8 years. Baron-Cohen further reports an elevated prevalence of autistic individuals in STEM-related careers \cite{baroncohen}. 
Finally, one study investigates the prevalence of individuals with ASD in computer science majors \cite{white_2011_college}. The authors find that  there is a relation between computer science majors and higher scores on the autism spectrum quotient measurement instrument \emph{AQ} \cite{baron2006autism}. They found that over 50 percent of the university students who scored high on the \emph{AQ} were pursuing computer science.
%Another study conducted in 2011 in the US, focused on \emph{HFA} among college students. This research included a sample of 685 students, where 667 participants had complete \emph{ASQ} data. The participants were undergraduate students enrolled in a large, technology-oriented, public university in the southeastern United States. Based on the data, they estimated that between 1 in 130 and 1 in 53 college students likely meet the criteria for \emph{HFA} \cite{white_2011_college}.

%To the best of our knowledge, no studies focus specifically on the prevalence of autism in the population of Software Engineering students and professionals. However, Stuurman et al. \cite{Sylvia} focus slightly on this topic but never give an estimate of the prevalence, as said in the study \textit{measuring the prevalence of autism among students is difficult. It is not possible, for obvious reasons, to diagnose each student as part of an investigation.}

\subsection{Neurodiversity in software engineering}
There is an increasing research interest in identifying the challenges and strengths of neurodivergent software engineers. 
In the following, we summarise existing work on those two areas.

\subsubsection{Challenges}
\label{subsubsec:challenges}
Neurodivergent individuals often experience difficulties in interpersonal communication, particularly with the interpretation of social cues and participating in collaborative activities, such as team meetings and discussions~\cite{Morris,Marquez2024IST,costello,Menezes2025SBSI}. However, the nature of these interpersonal communication challenges is bidirectional, often referred to as the \textit{``double empathy problem''}~\cite{milton2012double,milton2022double}. An increasing number of studies support the claim that neurotypical individuals also misread social situations with neurodivergent individuals~\cite{davis2021new,wadge2019communicative}. 
These communication challenges often lead to feelings of exclusion and misunderstanding, reinforcing negative biases against neurodivergent professionals, which ultimately damage their career progression and result in relatively higher unemployment rates~\cite{costello, davies2024career,hendricks_2010_employment,Menezes2025SBSI}. 

Autistic individuals often struggle with navigating job interviews and complex work environments that are not adapted to their needs. These individuals are statistically more likely to change jobs and earn less than their peers with equivalent qualifications and experience~\cite{hedley2018transition}. An inclusive work environment with adequate accommodations and support measures can significantly improve the performance and well-being of neurodivergent individuals~\cite{Gama2023Understanding,Marquez2024IST,Morris}. 
However, stigmata associated with autism often discourage individuals from disclosing their diagnosis, which in turn limits access to support~\cite{davies2024career, Menezes2025SBSI}. 

Neurodivergent individuals also face challenges concerning executive functioning and task management - especially those with ADHD - struggling more than neurotypical individuals in their ability to plan, organise, prioritise, and maintain focus, particularly when they need to switch between tasks~\cite{Gama2023Understanding,Gama2025ICSE-SEIS}. They are also prone to suffer from sensory and cognitive overload, as a result of distractions in the workplace -- a challenge in open-space offices -- as well as an overwhelming amount of work tools, both hindering concentration and impacting productivity~\cite{davies2024career,Marquez2024IST,Menezes2025SBSI}.

In the specific context of code reviews, communication challenges can lead to misunderstandings with the provided and obtained feedback. When giving feedback, a neurodivergent individual may struggle to find an adequate tone or may incorrectly interpret the tone of the received feedback~\cite{costello}. 
%Communication challenges can lead to stress and anxiety among neurodivergent individuals~\cite{Holmes2020dark}. 
Special attention to detail may lead to problems of losing focus on the broader context in a software artefact, which is detrimental to the code review process~\cite{Maun2023Participatory}. Finally, unstructured or ambiguous review processes can be challenging for neurodivergent individuals, as they feel more comfortable with well-defined processes and clear guidelines~\cite{Maun2023Participatory}.

\begin{figure}[ht]
    \centering
    \includegraphics[width=\linewidth]{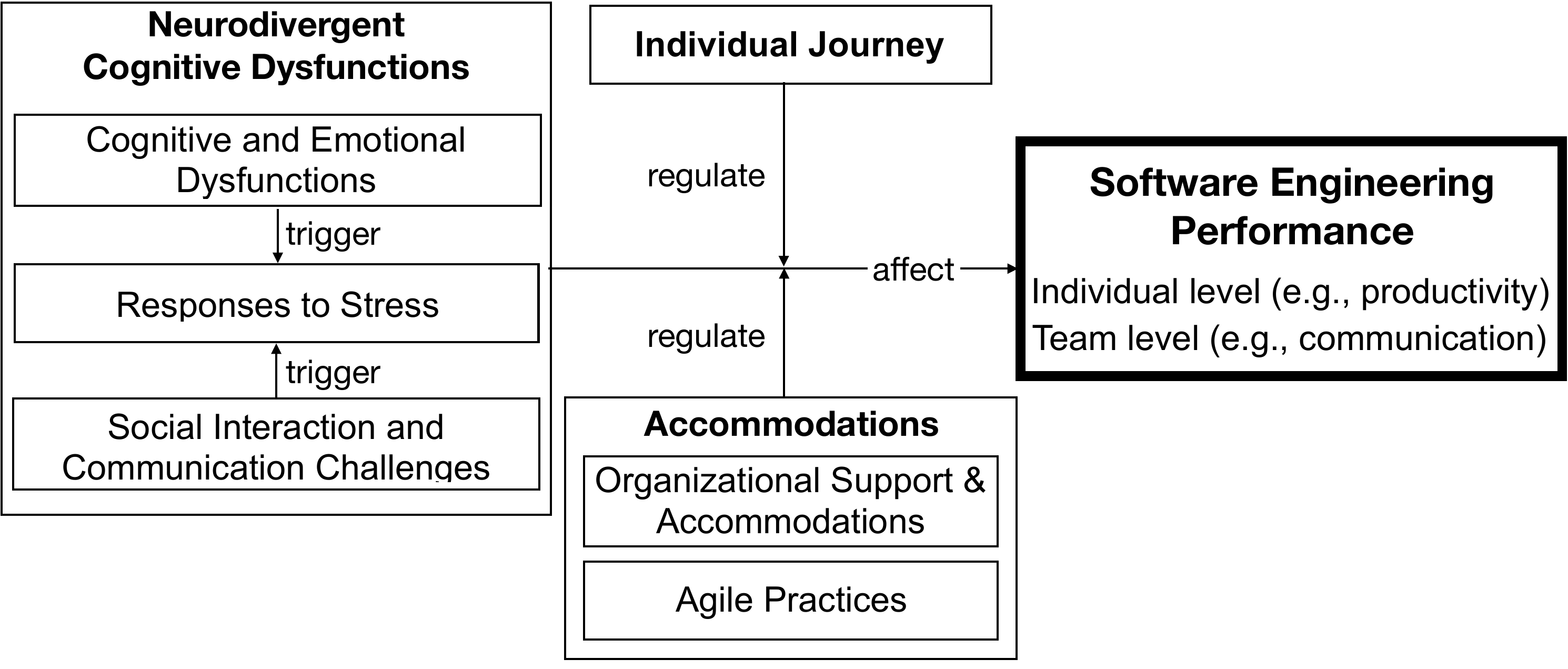}
    \caption{Theory on the Effect of Neurodivergent Cognitive Dysfunctions in Software Engineering Performance \cite{Gama2025ICSE-SEIS}. Reproduced with Authors' Permission.}
    \label{fig:theory_original}
    \Description{The figure shows Gama et al.'s original Theory on the Effect of Neurodivergent Cognitive Dysfunctions in Software Engineering Performance. The theory describes that Cognitive and Emotional Dysfunctions as well as Social interaction and Communication Challenges lead to a stress response. This, in turn, affects the individuals' software engineering performance, both on an individual and a team level. The individual's journey and existing accommodations moderate the effect. Within the accommodations, the theory distinguishes between Organisational Support and Accommodations and Agile Practices.}
\end{figure}

Gama et al.~\cite{Gama2025ICSE-SEIS} summarised their findings on difficulties of autistic and ADHD individuals in a theory, as depicted in Figure~\ref{fig:theory_original}.
This theory ultimately summarises the related work on challenges of neurodivergent individuals: it describes how various dysfunctions and communication difficulties lead to a stress response, which, in turn, affects their performance in SE tasks. Accommodations and the individual journay (diagnosis, acceptance, and other personal factors) can moderate this effect.

\subsubsection{Strengths}
As a part of natural diversity, neurodivergent individuals possess strengths in addition to their challenges.
Kirchner et al.~\cite{Kirchner2016Brief} identified open-mindedness, authenticity, love of learning, creativity, and fairness as the most prevalent strengths in autistic individuals. 
%An open-minded person likes to think things through, weighing the pros and cons, rather than jumping to conclusions, which can help in the process of performing a code review. 
Authenticity and fairness make an individual less likely to adapt their behaviour to improve their social reputation~\cite{izuma2011insensitivity}, and can lead to an increased focus on authenticity. 
Autistic individuals often pay attention to details~\cite{Marquez2024IST}, which, combined with a strong ability to focus, can help in detecting defects and inconsistencies in code~\cite{Annabi2017HICSS}.
The love of learning enables one to gain a deeper understanding of a particular topic of interest, such as a specific technology or codebase, making their insights extremely valuable in the context of a code review~\cite{Annabi2017HICSS}. Additionally, if the review topic aligns well with their interests, autistic individuals will likely have a strong motivation~\cite{Marquez2024IST} to perform a reliable and thorough revision of the software artefact. Creativity leads to original thinking and detailed elaboration~\cite{liu2011children}.

\section{Method}
%\grischa{Add Validity threats}
As we were investigating the experiences of autistic individuals in SE tasks, such as code reviews, and this is a socio-technical phenomenon, we used the STGT method~\cite{hoda_2022_sociotechnical}. We follow this up with a survey to lend further strength to the findings of the STGT. In the following sections, we discuss the details of this mixed-methods study.

\subsection{Socio-Technical Grounded Theory}
\subsubsection{Sampling and recruitment}
\label{sec:sampling_recruitmentSTGT}
We recruited 16 interviewees, all with an ASD diagnosis. 14 interviewees had a ASD level 1 diagnosis, of which five also had an ADHD diagnosis. Two participants were diagnosed with ASD level 2. One of those was diagnosed with Savant Syndrome. Most participants (13) received their ASD diagnosis at an adult age. Ten participants identified as men, while the remaining six identified as women.
11 participants were professionals and five graduate students, all with real (albeit, in students' case, short), recent, and relevant experience in conducting SE tasks, including programming and code reviews.
One participant has a post-graduate degree, seven an MSc degree, six a BSc degree, and the remaining two completed high school. 
We recruited participants from three countries (omitted here to preserve the double-anonymous review process). Apart from direct contacts ($n=2$), we reached out to ASD associations in multiple countries ($n=1$), as well as computer science departments ($n=1$). We also engaged with autism interest groups on social media platforms such as Discord, LinkedIn and Facebook ($n=3$). We also reached out to companies known for having special programmes for inclusion of autistic people - the most effective contact approach ($n=6$). Finally, we asked our participants to disseminate the study among their autistic contacts ($n=3$).
Table~\ref{table:stats} summarises key demographic data of the interviewees. We recruited one additional participant (I11), who later decided to drop out of the study.

\begin{table}[h!]
\caption{Demographic details of interview participants, including interviewee id (ID), their age range (Age), when they were diagnosed (\textbf{Diag.}), gender (Gen), current occupation, and highest achieved academic degree (Degree) -- High School (HS), BSc, MSc, or Post Graduate (PG).}
\centering
\begin{tabular}{ccccp{3.1cm}l} 
 \hline
 \textbf{ID} & \textbf{Age} & \textbf{Diag.} & \textbf{Gen} & \textbf{Occupation} & \textbf{Degree} \\ %[0.5ex] 
 \hline
 I01 & 30-39 & adult & W & Mechanical Eng. student & MSc \\ 
 %\hline
 I02 & 18-29 & adult & W & Web designer & MSc\\
 %\hline
 I03 & 40-49 & adult & M & IT industry & HS\\
 %\hline
 I04 & 40-49 & adult & M& Unemployed & BSc\\
 %\hline
 I05 & 40-49 & adult & M & Software developer & MSc\\ 
 %\hline
 I06 & 18-29 & child & W & Bio-informatics student & BSc\\
 %\hline
 I07 & 30-39 & adult & W & Comp. Science student & BSc\\
 %\hline
 I08 & 18-29 & child & M & Data engineer& MSc\\
 %\hline
 I09 &40-49 & adult & M& Unemployed & MSc\\
 %\hline
 I10 & 18-29 & child & M & Data engineer& PG\\
 %\hline
 I12 & 18-29 & adult&  M & Systems Analyst \& Dev & BSc\\
 %\hline
 I13 & 40-49& adult&  M & Electrician & BSc\\
 %\hline
 I14 & 30-39& adult& M  & Data Analyst & MSc\\
 %\hline
 I15 & 30-39& adult& M  & Linguistics Student & MSc\\
 %\hline
 I16 &18-29 &adult & W  & IT intern & HS\\
 %\hline
 I17 & 30-39 & adult & W & Data Analyst & BSc \\ [1ex] 
\hline
\end{tabular}
\label{table:stats}
\end{table}

%Out of the 16 participants, 14 individuals were diagnosed with ASD level 1, out of which 5 also had an Attention Deficit Hyperactivity Disorder (ADHD) diagnosis. Two participants were diagnosed with ASD level 2, among which one with Savant Syndrome. We did not have any participants with a diagnosis of ASD level 3.

%Our participants included seven from Portugal, eight from Brazil, and one from Poland. 

%To recruit participants, we reached out to multiple ASD associations in Portugal via email, as well as to computer science departments in universities from Portugal. Additionally, we actively made efforts to engage with ASD interest groups on social media platforms, such as \textit{Discord}, \textit{LinkedIn}, and \textit{Facebook}. Although the response from ASD associations was generally limited, we received the most positive feedback from a Facebook group.
%
%Specifically, we contacted \textit{Specialisterne}, a company that was founded with the mission include people with disabilities, particularly those with ASD. This contact yielded the majority of our participants.
%
%Finally, we asked participants to advertise our study in their circles.

\subsubsection{Data collection}
Participants signed an informed consent form prior to the interview.
Participation was voluntary and participants could withdraw from the study at any time.
Ethics approval was not required at the local institution.

All interviews were conducted online through Zoom, using a semi-structured interview guide.
We piloted the interview guide once prior to start of data collection.
We first asked about demographic information, including their ASD diagnosis, followed by questions on programming experience, code reviews, general challenges with ASD, and concluding the interview with questions on workplace culture and accommodations.
In line with STGT, the interview instrument was adapted throughout the study.
The final version of the interview guide is provided in this paper's companion site~\cite{companionSite}.
Interviews took between 17 and 49 minutes.

After conducting each interview, we immediately proceeded with the transcription and data cleaning process. 
Given that the majority of our interviews were conducted in \anon{Portuguese}, we encountered challenges in finding automatic transcription software compliant with privacy laws. 
Ultimately, we transcribed the interviews manually with the help of the tool oTranscribe\footnote{\url{https://otranscribe.com/}}.

\subsubsection{Data analysis}
\label{cha:data_analysis}
We opted for the \textbf{emergent mode} of theory development in STGT, which is used when studying a broad phenomenon \cite{hoda_2022_sociotechnical}. This mode of analysis leads to categories and indicative relationships of varying strength.
We iteratively conducted interviews, identified relevant pieces of data in the transcripts and coded them. We then used constant comparison of those codes to identify concepts that were then clustered into categories. Table~\ref{table:coding} shows two example categories and corresponding example quotes.
%\grischa{Add examples in a table?}\miguel{yes,created a base table for this}

\begin{table}[h!]
\caption{Coding evidence of two sub-categories that are part of the Adapting to change category}
\label{table:coding}
\begin{tabular}{|p{\linewidth}|}
\hline
\textbf{Category: Changing IDEs and other software tools}\\\hline
\#\textit{resistance} \\ ``with a lot of resistance and against my will'' - I03  \\\hline
\#\textit{adapting} \\ ``have some difficulty adapting to IDEs, but I solve it in a short time, a few hours, at most a day.'' - I10 \\
\hline \hline

\textbf{Category: Learning new languages}\\\hline
\#\textit{friction} \\ ``I don’t really like to start learning
anything new and I always feel a lot of
friction'' I01 \\\hline

\#\textit{adapting} \\ ``Change is never easy for me, but it
was a matter of adapting.'' I17 \\
\hline                                                 
\end{tabular}
\end{table}

We then graphically organised this information in the online whiteboard tool \textit{Miro}\footnote{\url{https://miro.com/}} using constant comparison, to identify key patterns in the data and to update emerging concepts and categories. 
Several categories emerged during this stage of analysis. However, no core category emerged, a phenomenon that is often anticipated in Grounded Theory methodologies.
In addition to coding, we wrote our reflections as memos next to the codes and categories that were starting to emerge, which led to relationships between codes and categories.

We followed this process until we reached theoretical saturation, which we measured by observing the amount of new concepts related to the research questions emerging after each interview.
After approximately 13 interviews, the amount of new categories dropped noticeably and did not increase again, so that we ended data collection and analysis after 17 interviews.

%\grischa{Should we mention theoretical structuring? It's unclear to me how it was used.}

\subsection{Validation survey}
\label{cha:evaluation_survey}
%\subsubsection{Sampling and recruitment}
After completing the STGT part of the study, we created and administered a survey to validate the grounded theory. 49 individuals completed the survey. Of those, only 5 reported having an ASD diagnosis, while the remaining 44 self-reported as neurotypical.
As a result, the survey answers give us a general picture of whether the observed categories apply (both to autistic and non-autistic individuals).
However, since the statistical power is low, statistical tests might be underpowered.

The survey was administered online through \textit{SosciSurvey}\footnote{\url{https://www.soscisurvey.de/}}. 
We employed a cross-sectional approach, where participants provide information at a single fixed point in time.
Opting for a self-administered questionnaire delivered online, we aimed to enhance accessibility and reach a broader audience. This method allows participants to respond at their convenience, aligning with the efficiency and inclusivity objectives of our survey \cite{shull_2008_guide}. 

The survey contained a brief introduction, followed by demographic questions and a section of questions related to the categories in our theory.
For this latter section, we opted for likert-scale agreement questions. Additionally, we incorporated residual options to prevent respondents from defaulting to the neutral position (Neither Agree nor Disagree) in the verbal anchor.
The questions in the survey were presented randomly to the participant, to ensure that participants encounter questions from different categories in a varied sequence.
To prevent the survey from becoming overly lengthy and to improve the likelihood of participant engagement, we opted to confirm only key points from each category.
%Our approach guarantees the representation of all categories within the survey, with each category featuring at least one related question.

We recruited autistic participants using the same sampling approach as before. We also reached out to participants from our interviews, inviting them to participate in the survey. Additionally, we extended invitations to individuals who had expressed interest in participating in interviews but did not respond to our follow-up emails. To recruit non-autistic individuals, we opted to distribute the survey within software engineering groups on social media platforms.
We obtained informed consent from participants and provided them with clear information on our data handling practices.
49 individuals completed the survey, of whom only 5 listed an ASD diagnosis. The participants had an average age of 29.5 years, with the oldest participant being 56 and the youngest 21 years old. 17 participants identified as women, 31 as men, 1 as non-binary. Considering the under-representation of women in SE, and the lower percentage of ASD diagnoses for women compared to men, the number of women is therefore quite high in this survey. The majority of participants are from 5 countries and 3 continents (anonymised here for the sake of the review process, but to be disclosed in the final version of the paper).
%from Portugal (31), followed by Belgium (6), USA (3), Brazil (3), and Canada (1).

To analyse the survey data, we employed \textit{Mann-Whitney U} tests \cite{mannwhitney} to compare the answers of the autistic and non-autistic participants. Throughout our analyses, we maintained a significance level of 0.05 for the \textit{p-value} to determine statistical significance.  %Additionally, we used \textit{Cronbach's} Alpha to quantify the internal consistency of questions relating to a single category~\cite{Cronbach1951Alpha}. 

\subsection{Validity threats and limitations}
\subsubsection{Construct validity}
During the interviews, we collected interviewees' self-reported experiences, which may introduce biases if they do not accurately recall or convey facts about their experiences as software engineers. 
Our survey was designed based on the qualitative results of the interviews. It is possible that some constructs were not fully captured in the interview process, and we tried to triangulate with the survey data to mitigate this threat, particularly with respect to the extent to which our findings represent concerns specific to autistic software engineers. As with the interviews, the survey also captures self-reported perceptions, which is a limitation. However, it is beyond the scope of this work to directly monitor software engineers to collect direct data from their work environment (e.g. efficiency and effectiveness measures of performance in coding and reviewing activities).

\subsubsection{Internal validity}
We followed the STGT process to build up our theory, interleaving interviews with coding activities. Researcher bias could potentially influence the qualitative coding of data. To mitigate this threat, the first author conducted initial coding of the qualitative data, followed by coding by the other authors. We discussed discrepancies and agreed on the coding scheme used in this paper.

\subsubsection{Researcher positionality} None of the authors is autistic. As we discuss in Section~\ref{subsubsec:challenges}, the so-called double empathy problem could potentially make communication harder during the interview process, as communication challenges are bi-directional. As we were aware of this threat, we were careful in our interactions with interviewees to mitigate it, as well as during our data interpretation. Additionally, we discussed early findings of this work with a fellow neurodivergent software engineering researcher to mitigate the risk of misinterpreting the data.

\subsubsection{Conclusion validity}
Our findings are limited by the sample sizes in the STGT and the survey, the latter with limitations in statistical power (e.g., a small number of group participants, particularly in the autistic group) and the potential for type I/II errors. We used appropriate statistical tests to mitigate this risk. However, the survey results in particular have to be interpreted with caution. 

\subsubsection{External validity}
The sample size of the STGT inquiry limits the generalisability of the results to the broader population of autistic software engineers. Autism is a spectrum, with many nuanced strengths and challenges among autistic software engineers, so the theoretical saturation may have a ``local'' nature (in the sense we may have been unable to recruit other autistic software engineers offering additional perspectives; or that questions might have been too limited to capture other, more important observations). As participation was entirely voluntary, there may also be a self-selection bias. We reached out to communities where we could find autistic software engineers, but it is possible that those who declined to participate would have offered alternative views in their interviews. We had interviews with participants from three countries and two continents, which provided some diversity, but having a more varied background could have also offered additional perspectives, broadening our theoretical insights. 

Concerning the validation survey, we used convenience sampling, drawing on our contacts, social networks, organisations supporting autistic people in general and software engineers in particular, and general communication channels in software engineering, because we wanted to attract both autistic and non-autistic respondents. The number of respondents (49) is not as large as we would have liked, particularly among autistic respondents, which may have impaired our ability to identify differences between autistic and non-autistic software engineers.

Our focus was essentially on programming and code reviewing during our interviews. Focusing on other software engineering activities could surface different opportunities and challenges.

Although this is an interdisciplinary research, the findings are restricted to autistic software engineers, potentially limiting their applicability to other technical domains and other categories of neurodiversity.

\section{Results}

%\begin{itemize}
%    \item Problem Solving
%    \item Self-efficacy in CR (and programming) / social interaction in CRs
%    \item Efficiency / Speed
%    \item Learning new things (tools, languages, ...)
%    \item Communicating through code - abstractions/unorganised code/logical thinking/effective comments (double empathy problem)
%    \item Communication modality (verbal feedback, written communication)
%    \item Flow / Focus
%    %\item Resume activities after interruption
%    \item detection obvious errors
%\end{itemize}

\begin{figure*}[ht]
    %\centering
    \centerline{\includegraphics[width=\textwidth]{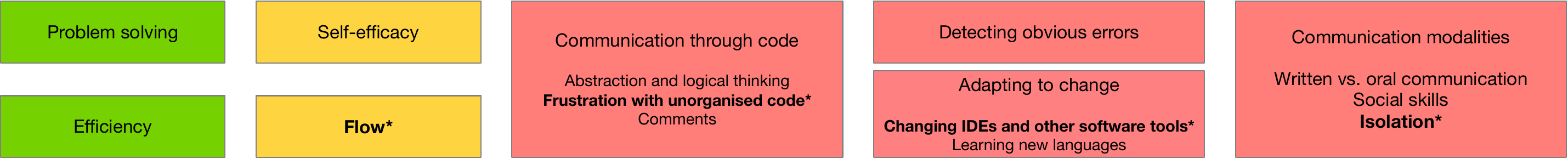}}
    \caption{Categories Overview. Green rectangles indicate a strength, red rectangles a challenge, and yellow rectangles a mixed strength/challenge category. Due to the amount of evidence, ``Communication through code'' and ``Adapting to change'' have sub-categories. Finally, bold-faced categories with a star (*) symbol indicate that at least some of the related questions in the follow-up survey showed a significant difference between autistic and non-autistic participants.}
    \Description{The figure shows the 8 categories. Problem-solving and Efficiency are both strengths. Self-efficacy and Flow are presented both as challenges and as strengths. Finally, Communication through code, Adapting to change, Communication modalities and Detecting obvious errors are the four challenge categories. Communication through code additionally has Abstraction, Frustration with unorganised code, and Learning new languages as sub-categories. Adapting to change has Changing IDEs and other software tools, and Learning new languages as sub-categories.}
    \label{fig:categories_overview}
\end{figure*}

In total, we identified eight categories, depicted in Figure~\ref{fig:categories_overview}. The colour coding indicates whether a category is a strength, a challenge, or a mixed strength/challenge category. In the following, we will present the categories in detail.

\subsection{Problem-solving}
Several interviewees self-reported strengths in problem-solving. \textit{``When the autistic person detects a problem, I speak for myself, he goes the second mile, and he goes beyond. This makes us able to solve problems more quickly, objectively, and assertively.''} I03. 
Additionally, some participants noted that they could see code differently, offering an interesting approach that set them apart. Others mentioned that their autism allowed them to spot patterns more clearly. \textit{``You can see the patterns of activities, it [autism] helps you at a certain point or even be aware of some detail of the code that is not making sense.''} I14. The emphasis on paying attention to small details was also reported to help catch defects effectively. \textit{``I can avoid the errors very well because I go into a lot of detail.''} I04.

We found no meaningful differences concerning the statement \textit{``I have very strong problem-solving capabilities.''} between autistic ($mean = 3.600, median = 4$) and non-autistic ($mean = 3.705, median = 4$) survey respondents.
While there seems to exist an underlying perception of persistence among our interviewees in solving problems or bugs, we found no statistically significant differences between non-autistic respondents in our survey. When asked about the level of agreement with the statement \textit{``If I encounter a problem or a bug in my code, I don’t rest until I can solve it.''}, autistic ($mean = 4.000, median = 4$) and non-autistic respondents ($mean = 3.636, median = 4$) lean towards being persistent in solving these issues. 
%Question 11 reveals distinct patterns between the autistic and non-autistic groups. Amongst the autistic group, unanimity is observed, with all participants either strongly agreeing, agreeing, or remaining neutral about their persistence in solving coding issues. The mean score of 4.0 indicates a strong inclination towards agreement as stated previously. In contrast, in the non-autistic group, approximately 48\% of participants agree with the statement, while around 21\% express disagreement, with an insignificant percentage strongly disagreeing. 
We also found no statistically significant difference in the agreement with the statement \textit{``I get stressed when I can’t solve a bug.''}, with autistic ($mean = 4.400, median = 4$) and non-autistic ($mean = 4.068, median = 4$) respondents.
%I get stressed when I can’t solve a bug. 
%In Question 30, both the autistic and neurotypical groups reveal strikingly similar trends in agreement with experiencing stress when unable to solve a bug. In fact, in the autism group, unanimity prevails, with no participants expressing neutrality or disagreement; instead, a significant majority (60\%) agrees, with an additional 40\% strongly agreeing with the statement.  Regarding the neurotypical group, we found that there is a minimal percentage (0.9\%) expressing disagreement or neutrality (0.4\%). The overall mean score is 4.102, so both groups are leaning toward agreement. 

\subsection{Self-efficacy}
When asked about situations where being autistic was an advantage in programming and code reviews, one interviewee expressed having a perception of a stronger ability than their peers. \textit{``I don’t know how to answer the question properly because I don’t know what normal is, I have no idea what normal is. So I’m going to assume that often being able to see patterns comes in handy, both in the data that we’re working with and how to process it. And logic, I feel like maybe understanding logic will be a little bit easier for me, but again, it’s unfair for me to be saying this without knowing what it is... this is normal for me.''} I10.

One interviewee emphasised the importance of code reviewing in boosting their self-confidence, \textit{e.g.}, through helpful discussions and feedback. Feeling understood and supported by their peers made the interviewee more comfortable and confident in their work.
However, some participants additionally shared a strong tendency for thoroughness, often exceeding requirements to ensure a comprehensive understanding of their tasks and to ensure they were well completed. While this commitment may initially appear to be a desirable trait in an employee, they reported sometimes feeling overwhelmed. This is derived from the challenge of determining when their work reached a satisfactory point, often resulting in a struggle to recognise when to stop. One participant reported that they would sometimes take a lot of time making their code perfect, and wonder if it is really good to be sent for review.

In our survey, we asked about the respondents' level of agreement with the statement \textit{``I need constant validation of my work.''} Although autistic respondents ($mean = 3.600, median = 4$) had a higher level of agreement than non-autistic respondents ($mean = 2.727, median = 2.5$), which could represent a lower perception of self-efficacy, the difference is not statistically significant. When asked about the agreement with the statement \textit{``Receiving feedback is good for me and it makes me feel confident about my work.''}, autistic ($mean = 4.000, median = 4$) and non-autistic ($mean = 4.159, median = 4$) respondents expressed a close level of agreement, without statistically significant differences. Concerning thoroughness, we asked about agreement with the statement \textit{``I pay close attention to detail.''} and found no statistically significant difference between autistic ($mean = 4.000, median = 4$) and non-autistic ($mean = 3.750, median = 4$) respondents. We complemented this with asking about the agreement with \textit{``I am a perfectionist.''}. Again, we found no statistically significant difference between autistic ($mean = 3.800, median = 4$) and non-autistic ($mean = 3.591, median = 4$) respondents. 

\subsection{Efficiency}
An interviewee expressed a sense of being faster than their peers in detecting patterns. \textit{``I would look at the data and very quickly I could tell if it made sense or not if we had to recalculate this. This is what I noticed when people said `That was quite fast, how did you notice that?', I don’t know if that’s normal or not, but for me, that’s perfectly normal.''} I10.

%An interviewee noted how their strong ability to memorise code increases their speed in programming tasks. \miguel{quote needed}

%A Mann-Whitney-U test indicated that the ability to memorise code was greater for autistic respondents (Mean = 3.8) than for neurotypical respondents (Mean = 2.545) $p = .0058$.

%While the better memory of \miguel{IXX} seems to help their efficiency, this increased ability to memorise code does not translate into a perception of higher speed in finding solutions for most participants in our survey. 

However, we found no statistically significant difference in agreement with the statement \textit{``I usually find the solution first than my peers.''} between autistic ($mean = 2.200, median = 2$) and non-autistic ($mean = 2.814, median = 3$) respondents, albeit the latter were slightly closer to a neutral answer, \textit{i.e.}, having a self-perception of being as fast as their peers, than autistic respondents.

\subsection{Adapting to change}
\subsubsection{Changing IDEs and other software tools}
6 interviewees reported struggling with adapting to new software tools, including Integrated Development Environments (IDE), as nuances in functionality and features differing from what they were accustomed to are a hindrance that makes them reluctant to change, expressing \textit{``difficulty in letting go of my way of doing things.''} I16.
For some, this is a significant barrier and changing an IDE is done \textit{``with a lot of resistance and against my will''} I03. Others deal with it easier. \textit{``[I] have some difficulty adapting to IDEs, but I solve it in a short time, a few hours, at most a day.''} I10. Changing IDEs and other tools is not problematic for all: 3 interviewees described a relatively swift adaptation process, reporting no difficulties when transitioning between \emph{IDEs} or learning new software. 

We further looked into this in our questionnaire. A Mann-Whitney-U test indicated that the agreement with the statement \textit{``I have trouble adapting to new tools and software.''} was greater for autistic respondents ($mean = 3.400, median = 3$) than for non-autistic respondents ($mean = 2.273, median = 2$),
$p = 0.0397$.

The challenge seems to be more about adapting to change than about actually using different IDEs. In our survey, we asked respondents the extent to which they agreed with \textit{``I have trouble using different IDEs.''} and found no statistically significant difference between the responses from autistic ($mean = 2.600, median = 2$) and non-autistic ($mean = 2.488, median = 2$) respondents. This suggests a slight tendency not to have trouble using different IDEs in both groups.

\subsubsection{Learning new languages}
Three of our interviewees expressed resistance to learning new things. This is more of a friction issue than an actual difficulty in learning. \textit{``I don’t really like to start learning anything new and I always feel a lot of friction [...] but that doesn’t mean I don’t learn.''} I01. 

In our survey, we asked participants about their level of agreement with the statement \textit{``I have difficulties learning new programming languages.''} We found no statistically significant differences between the responses from autistic ($mean = 2.200, median = 2$) and non-autistic participants ($mean = 2.227, median = 2$), in both cases suggesting members of both groups are not too challenged by learning new languages.

\subsection{Communicating through code}
\label{sec:comm_code}
\subsubsection{Abstraction and logical thinking}
Navigating abstraction within programming can be challenging for autistic software engineers. The abstract nature of programming concepts may represent heightened challenges due to differences in cognitive processing. Our interviewees' accounts illustrate how the complexity of abstract ideas can be especially perplexing. This struggle intensifies when deciphering abstractions created by others, as this requires an even deeper understanding of their thought patterns and cognitive structures. The cognitive divergence associated with autism can potentially add another layer of complexity to this process. Our interviewees also note that as projects evolve into higher levels of abstraction, the intricacy deepens, aggravating the challenge. 
\textit{``The biggest difficulty I find when programming is to do abstraction, the concepts are very abstract. If there is some abstraction and I have to understand the abstraction of the other person, then it is even more difficult. When I developed something more abstract, it got more complicated.''} I12.

%\subsubsection{Logical thinking}
Logical reasoning in programming can be challenging, as two interviewees reported. \textit{``[...] logical reasoning, [...] I think that’s the main challenge for me.''} I14. Logical thinking can also be challenging in code reviews. In code reviews, this challenge is particularly relevant, as reviewers must decipher the logic of unfamiliar code to effectively assess its functionality, efficiency, and adherence to best practices. Two interviewees reported struggling to understand others' code logic. \textit{``Difficulties usually are understanding the logic if it is from someone else.''} I17.

When asked about the extent to which they agree with the statement \textit{``I often feel difficulty understanding other people's code''}, both autistic ($mean = 3.600, median = 4$) and non-autistic ($mean = 3.068, median = 3$) respondents expressed a neutral leaning towards an agreement stance, with no statistically significant differences between the two groups.

\subsubsection{Frustration with unorganised code}
Two participants expressed concerns about the code's structure and organisation. Given that individuals with autism tend to adhere to rules and patterns, they conveyed feelings of frustration upon encountering unorganised code that did not align with the standards they were expecting. \textit{``If someone sends me a code to look at and the code is not organised in a way that is logical to me, the first thing I do is reorganise the whole code, which is extremely frustrating for me.''} I01.

A Mann-Whitney U test indicated that the level of agreement with the statement \textit{``I often get feelings of anxiety when the code is not formatted in a certain way.''} was significantly higher for autistic ($mean = 4.000, median = 4$) than non-autistic ($mean = 2.705, median = 2$) respondents, $p = 0.0319$.

%\grischa{I've added this part based on the old Comments section.}
\subsubsection{Comments}
Related to organising code, interviews highlighted issues revolving around code comments. Among the participants, a common challenge emerged: effectively commenting code and being frustrated by others commenting in the ``wrong'' way. Many struggled to strike the right balance in the number of comments to make. 
\textit{``The commented parts were simple and obvious but were not the right ones. It causes me discomfort, frustration even.''} I05.
\textit{``Another difficulty is if there is a lack
of commentary. It is very difficult to work
with lack of documentation.''} I08.
%Additionally, the interviewees highlighted that inappropriate comments were recognised as potential sources of discomfort and frustration. One participant went beyond what we expected and made a suggestion to enhance the code review experience, keeping in mind individuals on the Autism Spectrum. 
One interviewee suggested incorporating brief explanatory comments within different code blocks. These comments would help clarify the purpose and functionality of each specific piece of code.
\textit{``Better documentation of the code,
with a brief explanation in each block saying
what is intended to make.''} I05.
%
%I have difficulties commenting on the code.

Interestingly, in the survey, both groups disagree with the statement \textit{``I have difficulties commenting on the code.''}, with no statistically significant difference between autistic ($mean = 1.800, median = 2$) and non-autistic ($mean = 2.500, median = 2$) respondents. Both groups agree with the statement \textit{``Comments are very important.''}, again with a non-significant difference between autistic ($mean = 4.200, median = 5$) and non-autistic ($mean = 3.977, median = 4$) respondents.

%For Question 22 the overall mean score is 3.122, representing a sense of neutrality regarding feeling difficulty understanding other peoples' code. Within the autistic group, the mean is 3.6, and for the non-autistic group is 3.068. For the autistic group, we can see that 80\% agree with it, with the rest disagreeing. The standard deviation for this question is 0.79, suggesting that the responses were clustered closely around the mean. This reinforces the initial observation of minimal differences between the two groups.

\subsection{Communication modalities}
\subsubsection{Written vs. oral communication}
Communication was a recurring topic among the participants, often raised even outside questions directly related to it. Overall, interviewees showed a clear preference for written communication. 
They justified this preference by explaining that written messages allow them to refer back to the conversation if needed, unlike oral communication. 
\textit{``I prefer by text even because I can
read, I don’t know, it is easier for me to
communicate by text.''} I15.
However, two individuals showed a preference for oral communication when they were on the receiving end of the interaction. 
Interestingly, one participant additionally expressed a strong preference for oral communication with the support of visual aids, such as drawings: \textit{``The best way to communicate with an autistic person when they want them to program is with a drawing.''} I01.

We asked two questions in our survey to capture different nuances of this category, namely \textit{ease} and \textit{preference} for written or oral communication. The level of agreement with the statement \textit{``It is easier to use written communication rather than verbal communication.''} was higher for autistic ($mean = 3.800, median = 4$) than non-autistic ($mean = 2.659, median = 3$) respondents, albeit with no statistical significance. We observed a similar result concerning the level of agreement with the statement \textit{``I prefer to communicate with others by using written communication rather than verbal communication.''}, higher for autistic ($mean = 3.800, median = 4$) than for non-autistic ($mean = 2.659, median = 3$) respondents. Again, this difference was not statistically significant.

\subsubsection{Social skills}
Some participants reported feeling fearful of offending or hurting colleagues when providing feedback or engaging in conversations, especially when feedback involved soft skills. 
\textit{``I have a really hard time when the feedback is focused on a matter more of soft skill than a technical of skill.''} I16.
However, many of these concerns also centred around giving feedback during Code Reviews or discussing specific tasks.
One participant stated a preference for feedback that focuses primarily on technical aspects rather than on the colleague's social skills.
They believed that this approach would reduce the likelihood of accidentally causing harm or saying something inappropriate.
Interestingly, despite the difficulty of giving feedback, the interviewees expressed feeling more at ease when receiving it. 
\textit{``I always prefer to talk.''} I01.
One participant even mentioned that getting feedback was important as it boosted their confidence in their work.

In the survey, autistic participants had a higher percentage of agreement to the statement \textit{``I often fear offending my colleagues when speaking to them.''}
However, there was no significant difference between autistic and non-autistic participants on this question, with similar means in the autistic ($mean = 3.600, median = 3$) and non-autistic ($mean = 2.705, median = 2$) respondents. This was complemented by two questions concerning providing and receiving feedback. Confronted with the statement \textit{``I have trouble giving feedback about other people’s work.''}, both autistic ($mean = 2.200, median = 2$) and non-autistic ($mean = 2.750, median = 2$) respondents leaned toward disagreement, with no statistically significant difference. As for the statement \textit{``I have trouble receiving feedback about my work from other people.''}, autistic ($mean = 2.800, median = 3$) and non-autistic ($mean = 2.318, median = 2$) respondents lean toward disagreement, again with no statistically significant differences.

\subsubsection{Isolation}
Given their preference for written communication, four interviewees found pair programming challenging and complex. They often felt more pressure during these collaborative sessions. \textit{``I used to isolate myself a lot and I would rather be alone, even for hours on end, than have to deal with other people because my way of seeing things and theirs is completely different.''} I05.

An additional aspect of communication with peers is that most interviewees preferred working individually rather than in teams. In addition, they reported feeling comfortable working alongside artificial intelligence tools. Interestingly, a few even expressed a preference for interacting with AI chatbots over real people, stating that they felt more comfortable and relaxed.
\textit{``I get to have a healthier conversation with artificial intelligence than with
people. Today, my main contributor is ChatGPT.''} I03.

In the survey, we asked about the level of agreement with the statement \textit{``I prefer remote work rather than in person.''} A Mann-Whitney U test showed a higher level of agreement with this statement among autistic respondents ($mean = 4.600, median = 5$) than among non-autistic respondents ($mean = 3.163, median = 3$), $p = 0.0154$.
The level of agreement with the statement \textit{``I have trouble doing pair programming.''} was higher among autistic ($mean = 3.200, median = 4$) than non-autistic ($mean = 2.538, median = 2$) respondents.
However, the difference was not statistically significant. 
We also found no statistically significant difference in the level of agreement with the statement \textit{``I feel comfortable working with AI tools.''} Both autistic ($mean = 3.750, median = 4$) and non-autistic ($mean = 3.641, median = 4$) respondents expressed a slight tendency towards agreeing with the statement.

%Several participants additionally emphasised the importance of clarity in the information provided. They pointed out that clear instructions enable them to perform tasks more effectively. In contrast, overly abstract information can lead to confusion and hinder their ability to understand and complete tasks efficiently, as previously discussed.

%Assessing comfort with AI tools (Question 27), both groups exhibit a significant degree of comfort, particularly in the autistic group where a substantial majority (75\%) agrees with the statement and the rest presented neutrality. In the non-autistic group, while the majority also expresses comfort, there is a notable portion (46\% combined) that disagrees or is neutral. Additionally, there is 1 participant from the autistic group and 5 participants from the non-autistic group who did not provide any response.

\subsection{Flow}

As anticipated prior to our study, several interviewees highlighted topics related to focus, particularly in the context of programming. 
However, we name the category \textbf{flow}, as the phenomenon most interviewees described seemed closer to the flow concept known from existing work, i.e., a ``deep, focused, rewarding concentration'' \cite{mikkonen2016flow}.
That is, interviewees described themselves as good at maintaining concentration over longer periods and accomplishing tasks. For instance, participant I16 remarked:
\textit{We [autistic individuals] have the matter of hyperfocus, we start doing one thing, and if we focus too much on this thing, we end up producing a lot.}.
However, several interviewees also reported challenges alongside the benefits of flow.
One recurring theme was the difficulty they faced in diverting their attention away from a task once they were deeply engrossed in it. Moreover, some participants highlighted the struggle they encountered in disengaging from thoughts about a problem, even when they encountered it.
\textit{``I feel that if I have a problem, I don’t
know, I sometimes stay in that, I can’t stop
thinking about it.''} I15.

%focus issues
Some participants reported difficulty maintaining focus on tasks, particularly when instructions were unclear. This issue ties back to the discussion in Section~\ref{sec:comm_code} category, and clearly relates more to a flow state than focus.
\textit{``I consider myself someone who has a focus issue, and this is getting emphasised if the task is not well defined.''} I07.
However, we note that this difficulty might relate to existing co-morbidity with ADHD, as it is characteristic of inattentive ADHD.

In a similar direction, participants reported obstacles related to resuming to programming. That is, they encounter difficulties in re-establishing focus on programming tasks after interruptions or breaks. For instance, when they need to pause and then resume their work, they struggle with re-engaging their thought process and reasoning abilities. 
\textit{``My challenge may be when we’re doing the code, then we pause, and when I come back, I have to do the reasoning again.''} I02. Once again, this issue might be related more to ADHD than ASD.

%An interesting twist on this is that some respondents mentioned how they sometimes would temporarily step away from a problem.

%\grischa{Communicating through code or Flow?}
%Regarding challenges when coding, strategies such as stepping away from the task and returning with a renewed perspective resonate. Additionally, physical movement finds prominence as a means of alleviating stress. Participant I09 stated their difficulty when managing stress by saying: 

%\textit{It is complicated for me to manage anxiety when the code has many errors}

%I often experience hyperfocus when programming.
In the survey, a Mann-Whitney U test indicated that the level of agreement with the statement \textit{``I often experience hyperfocus when programming.''} was higher for autistic ($mean = 4.600, median = 5$) than for non-autistic ($mean = 3.634, median = 4$) respondents was statistically significant $p=0.0344$. 

Dealing with interruptions can be tricky, as they break focus, and software engineers then have to regain it to resume the task they were doing. When asked about the level of agreement with the statement \textit{``I often feel I can’t focus after someone interrupts me.''}, autistic respondents had a slightly higher level of agreement ($mean = 3.800, median = 4$) than non-autistic respondents ($mean = 3.182, median = 3$), but the difference is not statistically significant.

But what happens when one feels stressed about the activity they are working on? We did not find a statistically significant difference in the level of agreement with the statement \textit{``I voluntarily interrupt what I’m doing as a coping mechanism for stress management.''}, although autistic respondents had lower agreement with it ($mean = 2.600, median = 2$) than non-autistic respondents ($mean = 3.405, median = 4$).

Finally, we asked about the level of agreement with the statement \textit{``I see hyperfocus more as a hindrance than as an advantage.''}. We found no statistically significant difference between autistic ($mean = 2.600, median = 2$) and non-autistic ($mean = 2.000, median = 2$) respondents, with both groups leaning towards disagreement.

\subsection{Detecting obvious errors}
More closely associated with programming and error detection, some participants reported difficulty identifying syntax errors, such as missing commas or closing brackets. Furthermore, we observed a trend in which the majority of individuals who raised this concern were using an \emph{IDE} that either did not highlight these types of errors or were using a simple text editor.
%
%Challenges in detecting errors that seem obvious
Also, within the context of error detection, additional participants highlighted their struggles with identifying prominent errors. Interestingly, these minor mistakes were deemed more challenging due to a lack of attention. This observation seems directly at odds with the significant number of participants who reported a tendency toward hyperfocus during programming tasks. 
We also observe that some interviewees who had previously mentioned experiencing hyperfocus also reported struggling with this lack of attention.
These issues could relate to co-morbid ADHD, and not to autistic traits.
We did not ask a survey question for this particular category.
However, as we in hindsight noted that it conflicts with related work accounts, we decided to report it here regardless.

\section{Discussion}
%\begin{itemize}
    %\item Summarise/answer RQ
    %\item Compare to theory - propose a new one?
    %\item Problem solving vs. literature on problem solving: Overclaim? Do not only rely on strengths. Accommodations remain important.
%\end{itemize}

% \item[RQ] What strengths and challenges do autistic individuals exhibit in SE-related activities?
\subsection{RQ: What strengths and challenges do autistic individuals exhibit in SE-related activities?}
Through a STGT study, we investigated the RQ ``What strengths and challenges do autistic individuals exhibit in SE-related activities?''.
We find eight distinct categories.
Specifically, interviewees report strengths in problem solving, efficiency in programming and debugging, prolonged flow state when programming, as well as self-efficacy in programming. 
However, for many of these strengths, we found no statistically significant difference between autistic and non-autistic participants in our follow-up survey. These relatively small observed differences, visible both in the mean and median statistics for several questions, albeit non-significant, suggest we may have some small effect sizes that would require larger sample sizes to achieve statistical significance.
In terms of challenges, our interviewees reported challenges with adapting to change, e.g., changing IDEs and other software tools or learning new programming languages.
As expected from previous work, interviewees express a preference for written communication, especially when they are giving feedback.
Additionally, the interviewees noted that communication challenges also persist in code, e.g., understanding abstractions that others have created, frustrations with code that appears unorganised to them, or understanding others' logic.
While flow was primarily reported as a strength, several interviewees also reported it as challenging, especially re-focusing after interruptions.

\subsection{Implications to research}
Our findings have various implications to research and motivate directions for future work.
In the following, we first discuss notable observations we made in relation to existing work.
Then, we discuss how our work extends existing work on neurodiversity in SE.

We observe that many interviewees reported \textbf{strengths in problem solving}.
This facet is also commonly reported in related work, e.g., by Morris et al.~\cite{Morris} in terms of detecting patterns in code.
However, our survey showed no significant differences between autistic and non-autistic participants.
While this could be due to low statistical power, existing work on problem solving skills among autistic individuals indeed shows that they have comparably lower problem solving abilities than non-autistic individuals \cite{williams2014associations,constable2018problem,yakubova2017improving,polo2024comparison,minshew1994verbal}.
We see three potential explanations for this discrepancy.
First, individuals might overestimate their problem solving capabilities and indeed perform similarly or worse than their peers.
Second, there might be an over-representation of twice exceptional individuals among autistic software engineers.
Twice exceptionality refers to individuals that are exceptionally talented or gifted, but also have a learning or cognitive disorder \cite{baum2021twice}.
These individuals might be more likely to succeed in higher education and on the job market compared to autistic individuals without exceptional talent or gift.
These twice exceptional individuals might indeed have superior problem solving skills compared to their peers.
Third, autistic individuals might simply have a different way of addressing and solving problems.
Such a difference could surface in some cases as an advantage, but potentially also result in challenges, e.g., as discussed in Section~\ref{sec:comm_code}.

Also connecting to the previous point, we observe that many interviewees report \textbf{difficulties in understand abstractions, code or logical thinking of others}.
This point might indicate that the double empathy problem~\cite{milton2012double,milton2022double} extends to programming and abstraction in SE.
That is, that autistic software engineers have difficulties understanding the logical flow, abstractions used and code written by non-autistic software engineers, and vice versa.
We believe this point should be further studied in the realm of SE, as it could have implications on how to compose teams, for potential accommodations for autistic engineers, but also for educating non-autistic engineers and organisations as a whole about autism.
In particular, if our explanation is correct, autistic software engineers would not have problems understanding the code and abstractions of other autistic engineers.

In \textbf{comparison to existing work on autistic software engineers}, we can draw the following parallels.

First, several of our categories have been mentioned previously in related work.
For instance, the interviewees and follow-up survey by Morris et al.~\cite{Morris} show strengths in pattern detection among autistic individuals, which we believe is related to problem solving skills; preferences for textual communication; and hyperfocus (related to our flow category).
Similarly, Gama et al.~\cite{Gama2025ICSE-SEIS} describe hyperfocus (both in positive and negative terms), a preference for written communication, difficulties adapting to change, difficulties with pair programming for autistic individuals, as well as a tendency towards perfectionism and gold plating (related to our self-efficacy category).
Additionally, the authors describe difficulties in dealing with criticism in code reviews, which has connections to our category on communication through code.

Second, we only observe one contradiction with existing categories, i.e., Morris et al.~\cite{Morris} describe that autistic software engineers have the strength to detect errors quicker than their peers.
In our sample, this is reported as a weakness, i.e., individuals have difficulties spotting simple errors, such as overlooking missing brackets.
We hypothesise that this might relate to co-morbid ADHD.
However, even if that is true, we did not have anyone stating that detecting errors is a strength of theirs.

Third, several of our categories have not been reported in existing work.
While adapting of change has been reported with respect to routines, activities and habits, our interviewees report this challenge for tool and skill adoption.
This could represent an important factor in SE, as tool use is central and as tools and skills can change rapidly.
Most centrally, there is almost no coverage in existing work on what we coin communication through code.
Márquez et al.~\cite{Marquez2024IST} report that autistic software engineers have difficulties with computational thinking and state a need for developing this skill.
They additionally relate this challenge to (a lack of) problem solving capabilities.
We hypothesise that this might rather a difference in computational thinking, logic and abstraction instead of a deficit.

\begin{figure*}[htb]
    \centering
    \includegraphics[width=.7\textwidth]{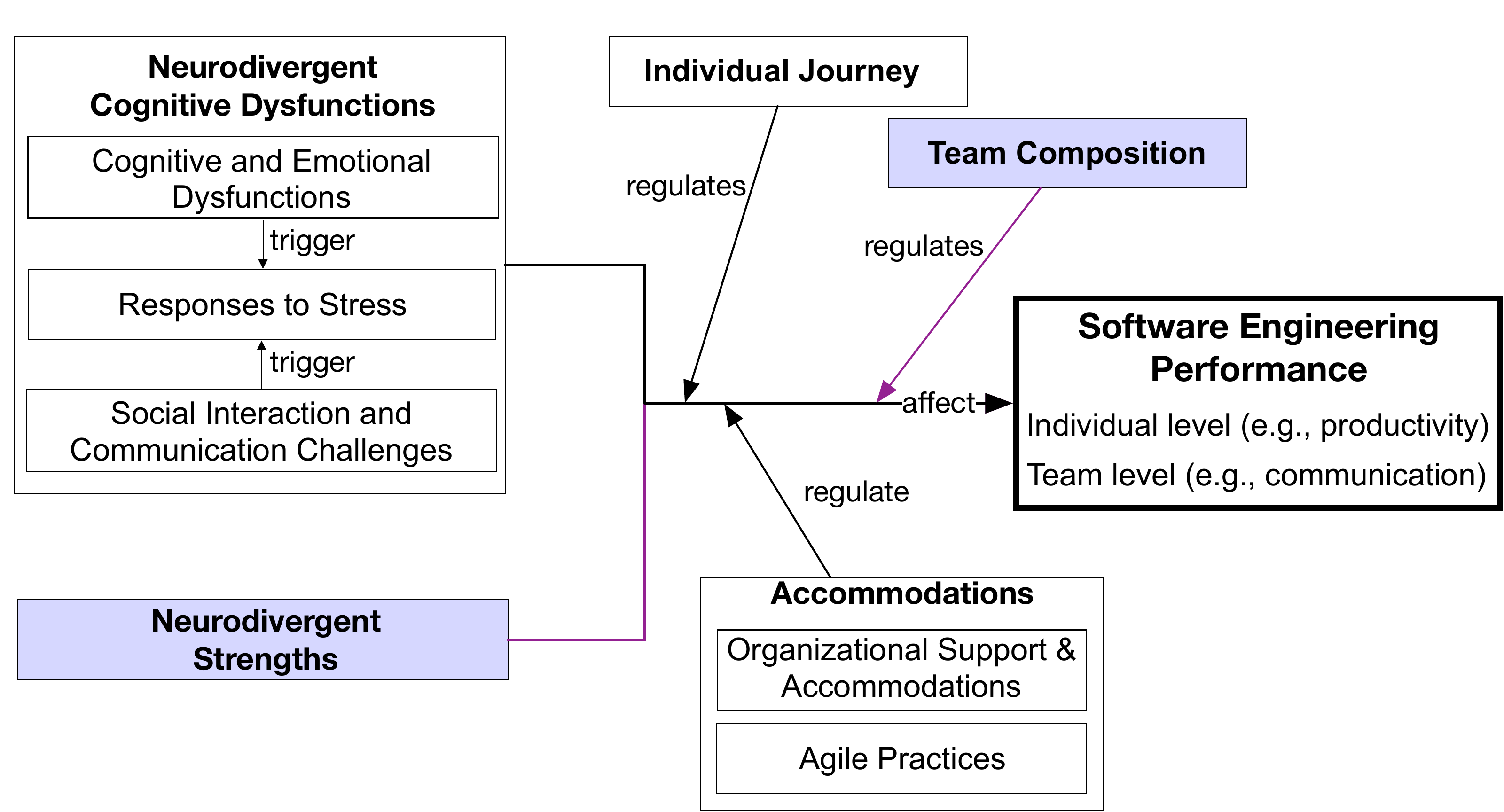}
    \caption{Updated Theory on the Effect of Neurodivergent Cognitive Dysfunctions in Software Engineering Performance. Additions are depicted in violet colour.}
    \label{fig:newTheory}
    \Description{The figure shows the extended Theory on the Effect of Neurodivergent Cognitive Dysfunctions in Software Engineering Performance. In addition to the original theory, we add that strengths of a neurodivergent individual can also affect their performance (usually positively). Additionally, we add that team composition also moderates the effect of both the dysfunctions and the strengths.}
\end{figure*}

Based on these comparisons to related work, \textbf{we propose extensions to Gama et al.'s~\cite{Gama2025ICSE-SEIS} Theory on the Effect of Neurodivergent Cognitive Dysfunctions in Software Engineering Performance}.
Our evidence supports the existing theory in substantial parts.
That is, our interviewees and survey participants describe how \textit{Cognitive and Emotional Dysfunctions} and \textit{Social Interaction and Communication Challenges} trigger \textit{Responses to Stress}.
Overall, this affects their performance in SE-related activities.
While not noted in the original theory, this effect is \textit{primarily negative}.
Finally, various \textit{Accommodations} and the \textit{Individual Journey} regulate the effect.
However, we also observe parts that are missing.
First, we note that \textit{individual strengths affect SE performance positively}.
These strengths are missing from the original theory.
Again, this effect is regulated by the individual journey and by accommodations.
For instance, despite many challenges, an individual might have learned to focus on their strengths, thus allowing them to leverage these better.
Similarly, accommodations (such as specific practices) might allow an individual to better make use of their strengths.
Second, in the original theory, the authors state that SE performance is affected on individual level and team level.
We hypothesise that, additionally, the team composition has an effect on an autistic individual's performance.
That is, challenges and strengths in problem solving, logical thinking, abstraction, and other cognitive activities might be heavily moderated by the team.
In fact, the same individual might experience their logical thinking as a challenge or as a strength, depending on the team context.
The resulting, updated theory is depicted in Figure~\ref{fig:newTheory}.

\section{Conclusion and future work}
In this paper, we reported on a Socio-Technical Grounded Theory study, investigating the strengths and challenges of autistic individuals in SE-related activities.
While previous work has studied autistic software engineers, it has primarily reported on general workplace challenges, such as working in open office spaces.
Instead, we focused on SE activities and included strengths in addition to the challenges.
We conducted 16 interviews with autistic SE professionals, and followed up with a survey among 49 participants (autistic and neurotypical).

Our findings reveal several strengths in problem solving, self-efficacy and efficiency in programming and other SE-related activities.
Additionally, we find novel insights regarding their challenges, e.g., that communication difficulties extend into code and abstractions.
We discuss our findings in relation to existing work, especially to the theory proposed by Gama et al.~\cite{Gama2025ICSE-SEIS}, and extend this theory with our novel findings.

Since our study added only minor additions to Gama et al.'s theory, we believe that a reasonable next step for future work is to validate the theory through quantitative studies, e.g., surveys or data collected from teams with autistic members.
Further work could investigate, in more depth, if and how the double empathy problem indeed applies to programming - since we only hypothesised this in the present study.
Finally, we believe there is ample opportunity to develop software tools designed to support neurodivergent individuals during code reviews and other SE-related tasks. These tools could be designed, e.g., to improve communication, feedback (pre-)processing and code comprehension, thereby boosting overall productivity and inclusivity.
In a similar direction, instead of tools, it might be feasible to adapt agile practices to make SE environments more inclusive to autistic individuals.

\newpage
\bibliographystyle{ACM-Reference-Format}

%\appendix



%Lorem ipsum dolor sit amet, consectetur adipiscing elit. Morbi
%malesuada, quam in pulvinar varius, metus nunc fermentum urna, id
%sollicitudin purus odio sit amet enim. Aliquam ullamcorper eu ipsum
%vel mollis. Curabitur quis dictum nisl. Phasellus vel semper risus, et
%lacinia dolor. Integer ultricies commodo sem nec semper.


%Etiam commodo feugiat nisl pulvinar pellentesque. Etiam auctor sodales
%ligula, non varius nibh pulvinar semper. Suspendisse nec lectus non
%ipsum convallis congue hendrerit vitae sapien. Donec at laoreet
%eros. Vivamus non purus placerat, scelerisque diam eu, cursus
%ante. Etiam aliquam tortor auctor efficitur mattis.


\end{document}